\documentstyle[11pt]{article}
\begin{document}
\begin{titlepage}
\centerline{\large\bf Cabibbo-Kobayashi-Maskawa Matrix}
\centerline{\large\bf Unitarity Triangle
                 and Geometry Origin of the Weak CP Phase}
\vspace{1cm}

\centerline{ Yong Liu }
\vspace{0.5cm}
\centerline{Lab of Numerical Study for Heliospheric Physics (LHP)}
\centerline{Chinese Academy of Sciences}
\centerline{P. O. Box 8701, Beijing 100080, P.R.China}
\vspace{0.2cm}
\centerline{E-mail: yongliu@ns.lhp.ac.cn}
\vspace{2cm}

\centerline{\bf Abstract}
\vspace{0.6cm}

In this work, the postulation that weak CP phase originates in
a certain geometry, is further discussed. According to this postulation,
the weak CP phase is determined by three mixing angles. So, if we
can determine experimentally three elements of the
Cabibbo-Kobayashi-Maskawa (CKM) matrix, we can then determine the
whole CKM matrix and correspondingly, the unitarity triangle.
We find that the angle $\gamma$ is about $\pi/2$ and the weak CP
phase $\delta\;(\delta_{13})$
only can exist in the first or fourth quadrant.
The conclusions coincide with the relevant analysis.
Some other predictions are given in this paper, the comparison of
the predictions based on the postulation
to the relevant experimental and theoretical results is listed.
All the predictions are consistent with the present
experimental results. 
\\\\
PACS number(s): 11.30.Er, 12.10.Ck, 13.25.+m

\end{titlepage}

\centerline{\large\bf Cabibbo-Kobayashi-Maskawa Matrix}
\centerline{\large\bf Unitarity Triangle
                 and Geometry Origin of the Weak CP Phase}
\vspace{1cm}

Quark mixing and CP violation [1-7] is one of the most interesting and
important problem in weak interaction and the whole particle physics.
Up to now, people do not know the real origin of CP violation. In the
standard model, CP violation originates from a phase presenting in the
CKM matrix [8-9]. Mathematically, a phase is permitted to present in
the three by three unitarity matrix. But, few people have asked the
question, what is the mathematical and physical meaning of the phase
and how is it generated ?

In the previous works [10], we have postulated that, the weak CP phase
originates in a certain geometry. Here, we discuss further this issue.
The central purpose is to make some predictions and contrast the
predictions with the experimental results. Then, we can see that, if
our postulation is reasonable.

To make this paper selfcontained, let us repeat the previous discussion
firstly.

\vspace{0.5cm}
\centerline{\bf 1. CKM matrix in KM parametrization and $SO(3)$ rotation}

There are many parametrization of the CKM matrix, such as the
standard one advocated by the Particle Data Group [10-11] and that
given by Wolfenstein [12] etc. However, the original parametrization
chosed by Kobayashi and Maskawa is more helpful to our understanding on
the problem, it is [9]
\begin{equation}
V_{KM}= \left (
\begin{array}{ccc}
   c_1 & -s_1c_3& -s_1s_3 \\
   s_1c_2 & c_1c_2c_3-s_2s_3e^{i\delta}& c_1c_2s_3+s_2c_3e^{i\delta}\\
   s_1s_2 & c_1s_2c_3+c_2s_3e^{i\delta}& c_1s_2s_3-c_2c_3e^{i\delta}
\end{array}
\right )
\end{equation}
with the standard notations $s_i=\sin\theta_i$ and $c_i=\cos\theta_i$.
Note that, it is just the phase $\delta$ in $V_{KM}$ violates CP symmetry.
And all the three angles $\theta_1, \theta_2, \theta_3$ can be taken to
lie in the first quadrant by adjusting quark field phases. In the following
discussions, we will fix the three angles in the first quadrant.

It is easy to find that, the above matrix can be decomposed into a product
of three Eulerian rotation matrice and one phase matrix [14].
\begin{equation}
V_{KM}= \left (
\begin{array}{ccc}
   1 & 0 & 0 \\
   0& c_2 &-s_2 \\
   0 & s_2 & c_2
\end{array}
\right )
\left (
\begin{array}{ccc}
   c_1 &-s_1 & 0 \\
   s_1& c_1 & 0 \\
   0 & 0 & 1
\end{array}
\right )
\left (
\begin{array}{ccc}
   1 &0& 0 \\
   0& 1& 0 \\
   0 & 0 & -e^{i\delta}
\end{array}
\right )
\left (
\begin{array}{ccc}
   1 &0& 0 \\
   0& c_3& s_3 \\
   0 & -s_3 & c_3
\end{array}
\right ).
\end{equation}

From the above equation, we can see that, the phase $\delta$ is
inserted into the CKM matrix some artificially. Although it is
permitted mathematically, but it is not so natural physically.

Eq.(2) can easily remind us such a fact: the $SO(3)$ rotation of
a vector. Let us describe this issue more detailed.
We begin with a special example which has been written into many
group theory textbooks. Suppose that vector $\vec{V}$
is located on $X-$axis and parallel to $Z-$axis. Now, we want to
move it to the $Z-$axis. There are infinite ways to do so. Here,
as a special example, we consider two of the most special ways.

1. Rotate $\vec{V}$ round $Z-$axis, after $\theta_1=\pi/2$, it is rotated
to $Y-$axis. Now, it is still parallel to $Z-$axis. Then, continue
to rotate it,
but this time, it is round $X-$axis. After $\theta_2=\pi/2$, it
is moved to $Z-$axis, but now, it is anti-parallel to $Y-$axis. We
denote it as $\vec{V_1}$.

2. Rotate $\vec{V}$ round $Y-$axis, after $\theta_3=\pi/2$, it is moved to
$Z-$axis directly, but it is anti-parallel to $X-$axis now. We denote
it as $\vec{V_2}$.

Note that, all of the movements of the vectors decribed above are
the parallel movements along
the geodesics. Now, we find that, starting out from the same vector
at the same point,
through the different two rotation ways, we obtain two different vectors
at the same point. The difference is only their direction. However, if
we rotate $\vec{V_1}$ round $Z-$axis (for more general case, it is the
normal direction of the point on which $\vec{V_1}$ and $\vec{V_2}$ stand),
after $\delta=\pi/2$, then, we get the same vector as $\vec{V_2}$.

From the special example, we can see that, the result of twice non-coaxial
rotations can not be achieved by one rotation. They are different by a
"phase" $\delta$. For more general case, $\delta$ is given by a simple
relation in spherical surface geometry
\begin{equation}
\sin\delta=\frac{(1+\cos\theta_1+\cos\theta_2+\cos\theta_3)\sqrt{
1-\cos^2\theta_1-\cos^2\theta_2-\cos^2\theta_3+
2 \cos\theta_1 \cos\theta_2 \cos\theta_3}}{
(1+\cos\theta_1)(1+\cos\theta_2)(1+\cos\theta_3)}.
\end{equation}
The geometry meaning of the above equation is evident. $\delta$ is the
solid angle enclosed by the three angles $\theta_1, \theta_2,
\theta_3$ standing
on a same point. Or, $\delta$ is the area to which the solid angle
corresponding on a unit spherical surface.

It is very important to
notice that, people have realized that the magnitude of CP violation
is closely related to a certain area more than ten years ago [5].

\vspace{0.5cm}
\centerline{\bf 2. Phase, geometry and the weak CP phase}

Due to Berry's famous work [15], the phase factor has aroused the
theoretical physicists a great interests in the past fifteen years.
People have realized that, the phase is closely related to a certain
geometry or symmetry [16-17]. For a non-trivial topology, the
presence of the phase factor is natural. In quantum mechanics, The well
known example is the Aharonov-Bohm effect [18].

To make the readers get to know with how we reached such a
postulation - weak CP phase as a geometry phase, let us recall a
simple fact in relativity.

Suppose that there are two observers $A$ and $B$, $A$ observes $B$,
$A$ gets the velocity $\vec{V}$ of $B$, $B$ observes $A$, $B$ gets
the velocity $\vec{U}$ of $A$. It is evident that,
$\vec{V}= - \vec{U}$, i.e. $\vec{V}$ anti-parallel to $\vec{U}$.
However, if the third observer $C$ presents,
and $A$ and $B$ observe each other not directly but through $C$, it
will not be the above case. Suppose $A$ observes $B$ through $C$, $A$
gets the velocity $\vec{V^\prime}$ of $B$, $B$ observes $A$ through
$C$, $B$ gets the velocity $\vec{U^\prime}$ of $A$. Now,
although $|\vec{V^\prime}|=|\vec{U^\prime}|$,
$\vec{V^\prime}\not=- \vec{U^\prime}$, i.e., $\vec{V^\prime}$ and
$\vec{U^\prime}$ are not parallel to each other. A angle presents
between these two veclocity vectors.

What can we learn from the above example?

First, the presence of the
angle is closely related to the presence of the third observer. This
is very similar to the case of quark mixing. If we only have two
generations of quark, we have no the weak CP phase, but, once we have
three generations of quark, the weak CP phase will present.
It is just this point stimulates us relating the weak CP phase to the
geometry phase.

Second, the more important issue we should realize is that, although
the space in which the three observers exist is flat, the velocity
space is hyperboloidal. Or in other words, it is a non-trivial
topology. In such geometric spaces, the presence of the phase is very
naturally [16][19-23].

Now, if we notice that mathematically, $SO(3)\simeq S^2$ or
$SU(2)/U(1) \simeq SO(3)\simeq S^2$, and if the quarks have a $SO(3)$
horizontal hidden symmetry, then the phase can present, and the
spherical geometry relation Eq.(3) can be obtained.

\vspace{0.5cm}
\centerline{\bf 3. Postulation and standard parametrization}

As described above, we have postulated the ad hoc relation. That
is: the three mixing angles $\theta_1, \theta_2, \theta_3$ and the weak
CP phase $\delta$ satisfy Eq.(3).

If we use the standard parametrization [10-11] instead of KM
parametrization Eq.(1), and correspondingly, we transform the
constraint Eq.(3) into the one expressed by $\delta_{13},\;\;
\theta_{12},\;\; \theta_{23}$ and $\theta_{13}$, then it will be
more convenient and clear for the following discussions.

The stardand parametrization is
\begin{equation}
V_{KM}= \left (
\begin{array}{ccc}
   c_{12} c_{13} & s_{12} c_{13}& s_{13} e^{-i \delta_{13}} \\
   -s_{12} c_{23}-c_{12} s_{23} s_{13} e^{i \delta_{13}} &
   c_{12} c_{23}-s_{12} s_{23} s_{13} e^{i \delta_{13}}    &
   s_{23} c_{13}\\
   s_{12} s_{23}-c_{12} c_{23} s_{13} e^{i \delta_{13}}  &
   -c_{12} s_{23}-s_{12} c_{23} s_{13} e^{i \delta_{13}} &
   c_{23} c_{13}
\end{array}
\right )
\end{equation}
with $c_{ij}=\cos\theta_{ij}$ and $s_{ij}=\sin\theta_{ij}$ for the
"generation" labels $i,j=1,2,3$. As the KM parametrization, the real
angles $\theta_{12}, \theta_{23}$ and $\theta_{13}$ can all be made to
lie in the first quadrant. The phase $\delta_{13}$ lies in the range
$0<\delta_{13}<2 \pi$. In following, we will also fix the three angles
$\theta_{12}, \theta_{23}$ and $\theta_{13}$ in the first quadrant.

The corresponding contraint on $\delta_{13}, \theta_{12}, \theta_{23}$
and $\theta_{13}$ - or, the expression of our postulation in this
parametrization is
\begin{equation}
\sin\delta_{13}=\frac{ (1+s_{12}+s_{23}+s_{13})
                       \sqrt{1-s_{12}^2-s_{23}^2-s_{13}^2+
                       2 s_{12} s_{23} s_{13}} }{(1+
                       s_{12}) (1+s_{23}) (1+s_{13})}
\end{equation}

If we note that the transformation relation between these two
parametrizations
$$
c_1=c_{12} c_{13} \;\;\;\;\;\; s_1=\sqrt{1-c_{12}^2 c_{13}^2}
$$
$$
c_3=\frac{s_{12} c_{13}}{\sqrt{1-c_{12}^2 c_{13}^2}}\;\;\;\;\;\;
s_3=\frac{s_{13}}{\sqrt{1-c_{12}^2 c_{13}^2}}
$$
$$
c_2=\sqrt{\frac{s_{12}^2 c_{23}^2+c_{12}^2 s_{23}^2 s_{13}^2+
2 c_{12} c_{23} s_{12} s_{23} s_{13} \cos\delta_{13}}{
1-c_{12}^2 c_{13}^2}}
$$
$$
s_2=\sqrt{\frac{s_{12}^2 s_{23}^2+c_{12}^2 c_{23}^2 s_{13}^2-
2 c_{12} c_{23} s_{12} s_{23} s_{13} \cos\delta_{13}}{
1-c_{12}^2 c_{13}^2}}
$$
\begin{equation}
\sin\delta=\frac{s_{23} c_{23}}{s_2 c_2} \sin\delta_{13}
\end{equation}
and
$$
s_{13}=s_1 s_3 \;\;\;\;\;\; c_{13}=\sqrt{1-s_1^2 s_3^2}
$$
$$
s_{12}=\frac{s_1 c_3}{\sqrt{1-s_1^2 s_3^2}}\;\;\;\;\;\;
c_{12}=\frac{c_1}{\sqrt{1-s_1^2 s_3^2}}
$$
$$
c_{23}=\sqrt{\frac{c_2^2 c_3^2+c_1^2 s_2^2 s_3^2-
2 c_1 c_2 c_3 s_2 s_3 \cos\delta}{
1-s_1^2 s_3^2}}
$$
$$
s_{23}=\sqrt{\frac{s_2^2 c_3^2+c_1^2 c_2^2 s_3^2+
2 c_1 c_2 c_3 s_2 s_3 \cos\delta}{
1-s_1^2 s_3^2}}
$$
\begin{equation}
\sin\delta_{13}=\frac{s_2 c_2}{s_{23} c_{23}} \sin\delta
\end{equation}
we will immediately find the following symmetry
between these two parametrization
$$
c_1\rightleftharpoons s_{13} \;\;
s_1\rightleftharpoons c_{13}\;\;
c_2\rightleftharpoons s_{23}\;\;
s_2\rightleftharpoons c_{23} \;\;
c_3\rightleftharpoons s_{12} \;\;
s_3\rightleftharpoons c_{12} \;\;
\delta \rightleftharpoons \delta_{13}
$$
$$
{\rm KM \;\;\; parametrization}\;\;\; \rightleftharpoons \;\;\;
{\rm Stardand \;\;\; parametrization}
$$
Then, we can get Eq.(5) easily.

\vspace{0.5cm}
\centerline{\bf 4. Predictions based on the postulation}

What can we extract from this postulation? And, how about the
correctness of the conclusions extracted from the postulation?

1. To make $\theta_1, \theta_2 $ and $\theta_3
\;\;(0<\theta_i<\pi/2, \;\;i=1,\; 2,\; 3)$ enclose a solid angle,
the following relation among them should be satisfied.
\begin{equation}
\theta_i+\theta_j \geq \theta_k  \;\;\;\;\;\;
(i\not=j\not=k\not=i=1,2,3)
\end{equation}
Comparing Eq.(5) with Eq.(3), we can find that, $\delta_{13}$ is the solid
angle enclosed by $(\pi/2 -\theta_{12}), (\pi/2-\theta_{23})$ and
$(\pi/2-\theta_{13})$. Hence, for the standard parametrization, the
following relation should hold.
\begin{equation}
(\frac{\pi}{2}-\theta_{ij})+(
\frac{\pi}{2}-\theta_{jk}) \geq (
\frac{\pi}{2}-\theta_{ki})  \;\;\;\;\;\;
(i\not=j\not=k\not=i=1,2,3. \;\;\;\theta_{ij}=\theta_{ji})
\end{equation}
It can be checked that, Eqs.(8, 9) are easily satisfied by the present
experimental data [11].

2. According to the geometry meaning of $\delta\;(\delta_{13})$,
it is the solid angle
enclosed by $\theta_1, \theta_2, \theta_3\;
(\frac{\pi}{2}-\theta_{12}, \frac{\pi}{2}-\theta_{23},
\frac{\pi}{2}-\theta_{13})$.
So, if $0<\theta_i<\pi/2 \;\;( 0<\theta_{ij}<\pi/2 \;\;
i\not=j=1, 2, 3)$ and $
\theta_i+\theta_j>\theta_k \;\;((\frac{\pi}{2}-\theta_{ij})+
(\frac{\pi}{2}-\theta_{jk})\geq (\frac{\pi}{2}-\theta_{ki})  \;\;\;
i\not=j\not=k\not=i=1,2,3 \;\;\;\;\theta_{ij}=\theta_{ji})
$
are satisfied, then, $\delta\;\; (\delta_{13})$ only
can lie in the first quadrant.
At most, you can take the solid angle $\delta\;(\delta_{13})$
as $4 \pi-\delta \;(4 \pi-\delta_{13})$. So, $2 \pi-\delta \;
(2 \pi-\delta_{13})$ in the fourth quadrant is also permitted.
If we notice that, there exists the
symmetry $\theta_i\rightleftharpoons -\theta_i$ in Eq.(3),
it is not difficult to understand this probability. Hence,

The second and third quadrants for $\delta\;\;(\delta_{13})$
are excluded thoroughly.

The recent analysis of Buras, Jamin and Lautenbacher [24] indicates
that, $\sin\delta_{13}$ likely lies in the first quadrant.

3. The estimate of the angle $\gamma$ in the unitarity triangles is about
$\pi/2$. Where, $\gamma$ with the other two angles $\alpha$ and
$\beta$ are definied as
\begin{equation}
\alpha \equiv arg(-\frac{V_{td} V_{tb}^*}{V_{ud} V_{ub}^*})\;\;\;\;\;\;\;\;
\beta \equiv arg(-\frac{V_{cd} V_{cb}^*}{V_{td} V_{tb}^*}) \;\;\;\;\;\;\;\;
\gamma \equiv arg(-\frac{V_{ud} V_{ub}^*}{V_{cd} V_{cb}^*})
\end{equation}

According to the geometric meaning of $\delta_{13}$, it is the solid
angle enclosed by $(\pi/2 -\theta_{12}), (\pi/2-\theta_{23})$ and
$(\pi/2-\theta_{13})$. The up-to-date experimental data [11]
tell us that, $s_{12}=0.217$ to 0.222, $s_{23}=0.036$ to 0.042,
and $s_{13}=0.0018$ to 0.0014. It means that, $\theta_{12},
\theta_{23}$ and $\theta_{13}$ are very small. Approximate to
the first order, we can take $\delta_{13}$ as the solid angle
enclosed by three right angles. So we get $\delta_{13}\sim \pi/2$.
On the other hand, based on the definition of $\gamma$ in Eq.(10)
and the form of standard parametrization Eq.(4),
it is evident that, $\gamma \sim \delta_{13}$. Finally, with no
detailed calculation, we have got to know that, $\gamma \sim \pi/2$.

This conclusion coincides very well with the relevant analysis [25-26].

4. For the case of more than three generations, the number of the
independent phases is also $(n-1) (n-2)/2$, where $n$ is the number
of the generation. According to the geometry meaning of the phase,
the number of the
independent phase is equal to the number of the triangles which we
can draw among $n$ points on a spherical surface with the areas of the
triangles are independent.

5. By use of the direct calculation result of the box diagram [27-28]
\begin{equation}
|\epsilon|=\frac{G_F^2 m_K f_K^2 B_K M_W^2}{\sqrt{2} (12 \pi^2)
 \Delta m_K} [
  \eta_1 S(x_c) I_{cc}+\eta_2 S(x_t) I_{tt}+2 \eta_3 S(x_c,x_t) I_{ct} ]
\end{equation}
with $I_{ij}=Im(V_{id}^* V_{is} V_{jd}^* V_{js})$,
$x_c=m_c^2/M_W^2 \;\; x_t=m_t^2/M_W^2$ and
\begin{equation}
S(x)=\frac{x}{4} [ 1+
     \frac{3-9 x}{(x-1)^2}+\frac{6 x^2 ln x}{(x-1)^3} ]
\end{equation}
\begin{equation}
\begin{array}{lll}
S(x,y)=&x y\{[\frac{1}{4}+\frac{3}{2 (1-y)}-
              \frac{3}{4 (1-y)^2} ] \frac{ln y}{y-x}+\\
&               [ \frac{1}{4}+\frac{3}{2 (1-x)}-
               \frac{3}{4 (1-x)^2} ] \frac{ ln x}{x-y}-
               \frac{3}{4 (1-x) (1-y)} \}
\end{array}
\end{equation}
and the experimental data on $\epsilon$ in $K^0-\overline{K^0}$ system,
substituting the contraint Eq.(3) or Eq.(5) into the CKM matrix,
we can further test our postulation. This has been done in our previous
works [10].

6. Rephasing invariant and the maximal CP violation. Substituting
Eq.(3) or Eq.(5) into Jarlskog invariant, we find that, the maximal
CP violation presents in $\alpha\simeq 71.0^0$, $\beta\simeq 90.2^0$
and $\gamma\simeq 18.8^0$ in triangle {\bf db}, and correspondingly,
Jarlskog invariant is about $0.038$ [29].

Now, we can say that, the conclusions extracted from our postulation
all are consistent with the present experimental and theoretical
results. Perhaps, it is the continuation which has not been finished
thoroughly by Profs. Kobayashi and Maskawa in their original work.

\vspace{0.5cm}
\centerline{\bf 5. Conclusions and discussions}

In this paper, after summarizing our previous works, some new results
are given. They are

1. Transform the constraint Eq.(3) in KM parametrization into that in
stardand parametrization, it is the Eq.(5).

2. Predict undoubtedly that, if all the mixing angles are made to
lie in the first quadrant, the second and third quadrant for
$\delta\;(\delta_{13})$ are excluded thoroughly.

3. Predict that, the angle $\gamma$ is about $\pi/2$.

Furthermore, the comparison of the predictions based on our postulation
to the relevant experimental and theoretical results is listed. We find
that all the predictions coincide with the present experimental results.

Our postulation can be further put to the more precise
tests in $B-$factory in near future. If it can be verified finally,
it means that, only three elements in KM matrix are independent,
and hence we have removed one of the free parameters in the standard
model, the number of the free parameters is now eighteen other than
ninteen. If then, we will feel that the physics is more simple, natural,
and beautiful. But, we will naturally ask, what is the dynamic origin?

However, our postulation is only supported by the present experiments.
It is a ad hoc supposition now, it still need the further verification
by the future experiments and a theory on the more basic level on which
it can base.

Because the CP phase can originate from many ways in different theories
and physical processes, in the standard model, we hope that our
postulation at least provide part of the CP phase. Anyway,
even if our postulation has no any physical base, it is still a good
parametrization for the weak CP phase.

\end{document}